\documentclass[aps,prl,twocolumn,superscriptaddress]{revtex4-1}
\usepackage[pdftex]{graphicx}
\def \etal {{\textit {et al.}\ }}
\def \urs {URu$_2$Si$_2$}
\def \ursd {URu$_{2-x}$(Fe,Os)$_{x}$Si$_{2}$}

\begin{document}

%\preprint{}

\title{Electrodynamics of the antiferromagnetic phase in \urs}



\author{Jesse S. Hall}
\affiliation{Department of Physics and Astronomy, McMaster University, Hamilton, ON, Canada L8S 4M1}

\author{ M. Rahimi Movassagh}
\affiliation{Department of Physics and Astronomy, McMaster University, Hamilton, ON, Canada L8S 4M1}

\author{M. N. Wilson}
\affiliation{Department of Physics and Astronomy, McMaster University, Hamilton, ON, Canada L8S 4M1}

\author{G. M. Luke}
\affiliation{Department of Physics and Astronomy, McMaster University, Hamilton, ON, Canada L8S 4M1}
\affiliation{Canadian Institute for Advanced Research, Toronto, ON, Canada M5G 1Z8}

\author{N. Kanchanavatee}
\affiliation{Department of Physics, University of California, San Diego, La Jolla, California 92093, USA}
\affiliation{Center for Advanced Nanoscience, University of California, San Diego, La Jolla, California 92093, USA}

\author{K. Huang}
\thanks{Present address: State Key Laboratory of Surface Physics, Department of Physics, Fudan University, Shanghai 200433, China}
\affiliation{Department of Physics, University of California, San Diego, La Jolla, California 92093, USA}
\affiliation{Materials Science and Engineering Program, University of California, San Diego, La Jolla, California 92093, USA}

\author{M. Janoschek}
\thanks{Present address:  MPA-CMMS, Los Alamos National Laboratory, Los Alamos, New Mexico 87545, USA}
\affiliation{Department of Physics, University of California, San Diego, La Jolla, California 92093, USA}
\affiliation{Center for Advanced Nanoscience, University of California, San Diego, La Jolla, California 92093, USA}

\author{M. B. Maple}
\affiliation{Department of Physics, University of California, San Diego, La Jolla, California 92093, USA}
\affiliation{Center for Advanced Nanoscience, University of California, San Diego, La Jolla, California 92093, USA}
\affiliation{Materials Science and Engineering Program, University of California, San Diego, La Jolla, California 92093, USA}

\author{T. Timusk}
\affiliation{Department of Physics and Astronomy, McMaster University, Hamilton, ON, Canada L8S 4M1}
\affiliation{Canadian Institute for Advanced Research, Toronto, ON, Canada M5G 1Z8}
 

\date{\today}

\begin{abstract}

We present data on the optical conductivity of \ursd. While the parent material \urs\ enters the enigmatic hidden order phase below 17.5 K, an antiferromagnetic phase is induced by the substitution of Fe or Os onto the Ru sites. We find that both the HO and the AFM phases exhibit an identical gap structure that is characterized by a loss of conductivity below the gap energy with spectral weight transferred to a narrow frequency region just above the gap, the typical optical signature of a density wave. The AFM phase is marked by strong increases in both transition temperature and the energy of the gap associated with the transition. In the normal phase just above the transition the optical scattering rate varies as $\omega^2$. We find that in both the HO and the AFM phases, our data are consistent with elastic resonant scattering of a Fermi liquid. This indicates that the appearance of a coherent state is a necessary condition for either ordered phase to emerge. Our measurements favor models in which the HO and the AFM phases are driven by the common physics of a nesting-induced density-wave-gap. \end{abstract}

%\keywords{Hidden order, antiferromagnetism, nesting, Fermi liquid, optical conductivity}

\maketitle

%\section{}
%\subsection{}
%\subsubsection{}

The heavy fermion metal \urs\ has a rich phase diagram in both temperature and pressure \cite{amitsuka99,bourdarot05,matsuda03}. Uniquely among heavy fermion materials, as the temperature is lowered, the development of the heavy fermion phase is interrupted by a second order phase transition  at 17.5 K \cite{palstra86, maple86} to an enigmatic ``hidden order'' (HO) whose physical origin has been the subject of considerable study. Despite intense experimental investigation \cite{wiebe07, santandersyro09,schmidt10, aynajian10,hassinger10} and proposed theoretical models \cite{haule09,dubi11,chandra13, riseborough12,das14}, the nature of the phase transition has remained elusive. 

When hydrostatic pressure is applied to \urs\ the temperature of the transition rises steadily with pressure up to 20 K at 1.5 GPa, \cite{mcelfresh87} at which point a first order phase transition from the HO phase with a small extrinsic magnetic moment \cite{matsuda03,niklowitz10}  to a large moment long-range antiferromagnetic (AFM) phase \cite{amitsuka99,amitsuka07,bourdarot14} occurs. As the pressure is further increased the transition temperature continues to rise up to nearly 30 K \cite{hassinger08}. Much attention has been placed upon the antiferromagnetic phase because quantum oscillation measurements \cite{nakashima03,hassinger10}  suggest that the Fermi surface does not change between the HO and AFM phases. This allows calculations of the bandstructure \cite{elgazzar09, oppeneer11} and the Fermi surface, which can be computed for the AFM phase, to be applied to the HO phase.

Recently \cite{kanchanavatee11,das15,wilson15}, it has been found that the partial substitution of Fe onto the Ru sites in \urs\ can also induce antiferromagnetism in the system. Increasing the concentration of Fe increases the transition temperature and, as with applied hydrostatic pressure, there is a crossover into antiferromagnetism above a certain substitution level. The similarity of the phase diagrams with pressure and Fe substitution suggests that the AFM phases are equivalent. Substitution with Os also induces an AFM phase that raises the transition temperature \cite{kanchanavatee14, wilson15}. In this paper, we present optical conductivity data on \urs\  in the substitution-induced antiferromagnetic phase and we report the first observations of the behavior of the energy gap for charge excitations in the AFM phase and its evolution with increasing substitution $x$.  We also show spectra of the normal paramagnetic phase at temperatures above the AFM transition and contrast them with the spectra above the HO transition.

The Fe substituted crystals were grown in a tetra-arc furnace in San Diego and the Os substituted samples in a tri-arc furnace at McMaster, both using the Czochralski method in an argon atmosphere. Magnetic transitions were characterized using a SQUID magnetometer and the presence of an ordered antiferromagnetic moment was confirmed separately using muon spin rotation \cite{wilson15} and neutron scattering \cite{das15}. DC resistivity measurements were performed in an Oxford Maglab system using a four-probe geometry; the estimated error due to sample configuration and geometry is 20$\%$, which is consistent with the variation in measurements in the literature \cite{palstra86, jeffries07}. Optical measurements were performed using an SPS 200 Martin-Puplett Fourier-transform interferometer for reflectance measurements below 20 meV and a Bruker IFS 66v/s FTIR spectrometer for measurements from 15 meV to 4.5 eV. Absolute reflectance was measured using a standard gold evaporation technique \cite{homes93}, and the optical conductivity was obtained by performing a Kramers-Kronig analysis on the absolute reflectance data. 

\begin{figure}
\includegraphics[width=3.5 in]{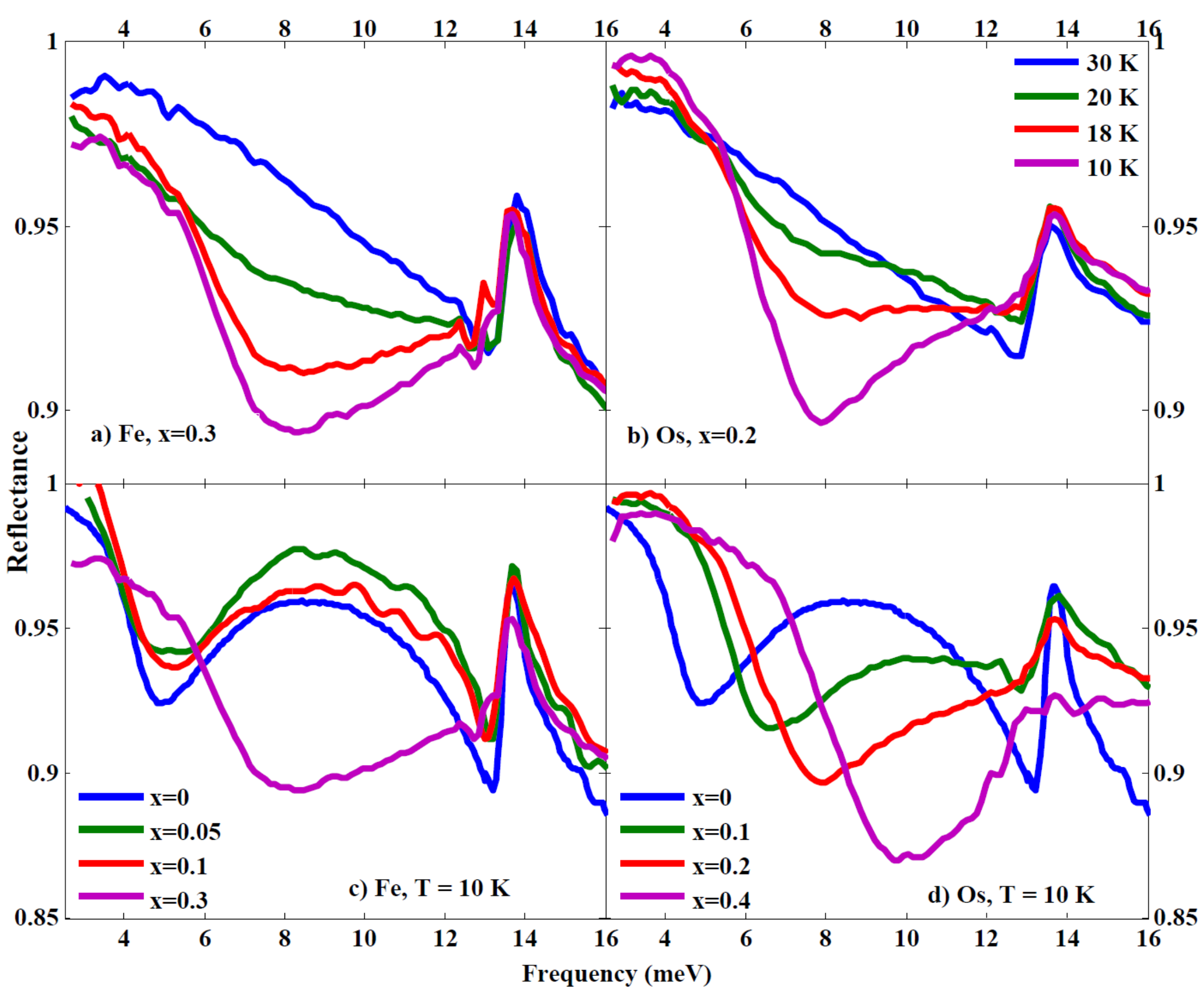}%
\caption{\label{reflectance}Temperature and substitution dependence of the reflectance of samples of \ursd. The top panels shows the absolute reflectance for a) Fe, b) Os substitution, both in the AFM phase as a function of temperature. The bottom panels show the absolute reflectance as a function of $x$ for c) Fe and d) Os substitution. The prominent depression of reflectance that develops in the 5 to 10 meV region in all the samples is a signature of a gap in the density of phases. Adding Fe and Os causes the reflectance minimum  to move to higher frequency but the signature of the gap, a single minimum of reflectance, does not change with substitution. }
\end{figure}
% change the rubrics, they do not agree with the text
%I will change the figure, it doesn't need to have the relative reflectance in it.

Figure \ref{reflectance} shows the absolute reflectance of \ursd\  in the AFM phase (panels a and b) and at different concentrations $x$ for Fe substitution (panel c) and for Os (panel d). The parent compound with $x=0$, which is in the HO phase, is shown as well. In all the curves a single strong minimum develops as the temperature is lowered below the transition. As $x$ is increased the reflectance minimum moves to higher frequencies.

The characteristic absorption that signifies the opening of the HO gap in the parent compound is still present in the AFM phase, remarkably unchanged except for a shift to higher energies. There is no evidence of a second, different gap due the AFM phase. Previously \cite{hall12} we have shown that when two gaps are present in \urs, as in the case of the c-axis conductivity, it is possible to see the characteristic change in the absorption due to this effect if the gap energies differ sufficiently. The absence of a second gap here strongly argues for a common gap-forming mechanism in the HO and AFM phases. 

\begin{figure}
\includegraphics[width=3.5 in]{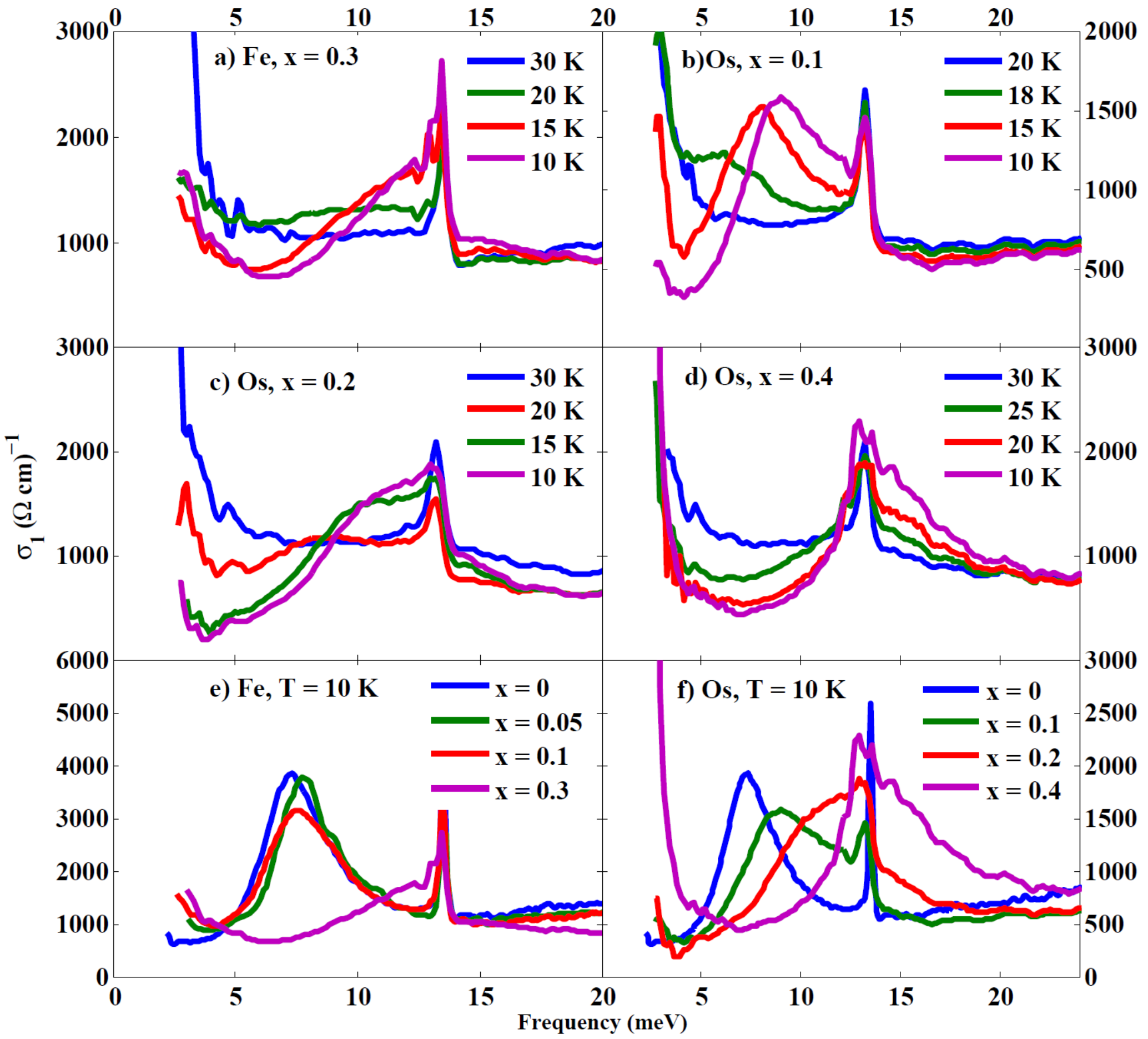}%
\caption{\label{conductivity}Optical conductivity of \ursd \ in the antiferromagnetic phase. Panels a) to d) show the optical conductivity changes with temperature for:  a) Fe x=0.3, b) Os x=0.1, c) Os x=0.2, and d) Os x=0.4. Panel e) shows the dependence of the conductivity on the concentration for Fe substitution, with the parent material shown for comparison; panel f) shows the same for Os substitution. The parent compound conductivities have been reduced by a factor of 0.5 to allow easier comparison. The sharp peak at 13.6 meV is an optical phonon.}
\end{figure}

Figure \ref{conductivity} shows the AC conductivity of \ursd. In the paramagnetic phase, above the HO and AFM transition temperatures, the conductivity consists of a Drude peak and an incoherent continuum. As in the unsubstituted sample \cite{hall12}, the continuum develops a gap-like minimum at the phase transition. This minimum is unaltered in overall character between the HO and the AFM phases: both show a characteristic depletion of spectral weight in the gap region followed by a recovery in the frequency range immediately above the gap. The Drude peak that develops in the hybridization regime narrows but is otherwise unaffected by the emergence of the ordered phases. The principal effect of Os and Fe substitution is an increase in the energy of the gap, in tandem with the increase in transition temperature.  

% introduce the table here with the appropriate caption
% would a graph of the gap be more appropriate? perhaps transition temperature on the x-axis and gap energy on y-axis?

We can characterize the gap in the electronic density of states at the Fermi level $\Delta$ using the method described in reference \cite{hall12} for the parent material, giving a reasonable estimate for the size of the gap and its temperature evolution. Results of this analysis are shown in Table 1. While the absolute value for the energy gap is somewhat model-dependent, we estimate the relative accuracy of the gap values between samples to be $\pm0.2$ meV. The ratio $2\Delta/k_BT_0$, where $T_0$ is the transition temperature to the ordered phase (HO or AFM), has two values, a lower value of 4.2 for the parent compound and small Fe substitution and a somewhat higher value of 5.2 in the more heavily-substituted AFM phase.  The charge gap closely tracks the transition temperature in both the AFM and HO phases, regardless of whether Fe or Os is used to induce antiferromagnetism. In particular, the value of 2$\Delta/k_B T_0$ has nearly the same value deep in the AFM phase for both Fe and Os substitution. This implies that the AFM phase is the same for all samples and is not specific to Fe or Os, in agreement with $\mu$SR studies of the two systems \cite{wilson15}. 

It is worth comparing these results to those obtained by other techniques in the antiferromagnetic phase. Resistivity measurements on \urs \ performed under pressure by Jeffries \etal \cite{jeffries08} and specific heat measurements with Fe substitution by Das \etal \cite{das15} suggest that, as the transition temperature rises, the gap in the electronic excitations remains constant in the HO phase, then rises to higher values in the AFM phase. In contrast, as the table shows, our measurements clearly demonstrate that the gap tracks the rising transition temperature in both the HO and AFM phases. 

\begin{table}
\begin{tabular}{l l l l l} \hline
Fe                   & $\Delta$ (meV) &$T_0$ (K) & $2\Delta/k_BT_0$ &phase  \cr \hline
x = 0               &     3.2             & 17.5       &  4.2     & HO                  \cr
x = 0.05          &     3.3           & 18.2       &  4.2     & HO/AFM                     \cr
x = 0.1            &     3.4           & 18.5       &  4.3     & HO/AFM                    \cr 
x = 0.3            &     5.1           & 23          &  5.2     & AFM                 \cr \hline
Os  & $\Delta$ (meV) &$T_0$ (K) & $2\Delta/k_BT_0$ &phase \cr \hline
x = 0               &     3.2             & 17.5       &  4.2    &HO                   \cr 
x = 0.1               &    4.4            & 19.5       &  5.2 &AFM                      \cr
x = 0.2               &     5.1            & 23       &  5.2     &AFM                  \cr
x = 0.3               &     6.6           & 29       &  5.3      &AFM                 \cr \hline
\end{tabular}
\caption{Gap values for the various levels of substitution by Fe and Os. The gap and the critical temperature increase monotonically in tandem with substitution. T$_0$ was determined from both resistivity and SQUID magnetization measurements. The phase of the x=0.05 and x=0.1 samples with Fe substitution may be a mix of HO and AFM (see reference \cite{das15} for a discussion of the phase diagram).}
\end{table}

We now turn to the excitations in the paramagnetic normal state above the ordered HO and AFM phases. Below 70 K, the conductivity in the 5 meV to 40 meV region decreases monotonically with decreasing temperature forming the so-called ``hybridization gap" \cite{nagel12}. The spectral weight lost in the hybridization gap is transferred to much higher frequencies \cite{guo12, hall14} while in contrast both HO and AFM phases shift their spectral to a new hump feature {\it immediately} above the gap. Another common feature of the HO and AFM phases is the ``arrested hybridization". In both phases the hybridization gap stops changing below the HO/AFM transition temperature. 

%\begin{figure}
%\includegraphics[width=3.5 in]{scattering_rate}%
%\caption{\label{scattering_rate} Scattering rate of the parent compound is linear with frequency above 30 K but develops an $\omega^2$ dependence %below a temperature where coherence sets in (dashed line) \cite{nagel12}.  We note that in the substituted compound the Fermi liquid region has moved %up in temperature to 30 K as shown in Fig.\ref{rho_ac}. }
%\end{figure}

\begin{figure}
\includegraphics[width=3.5 in]{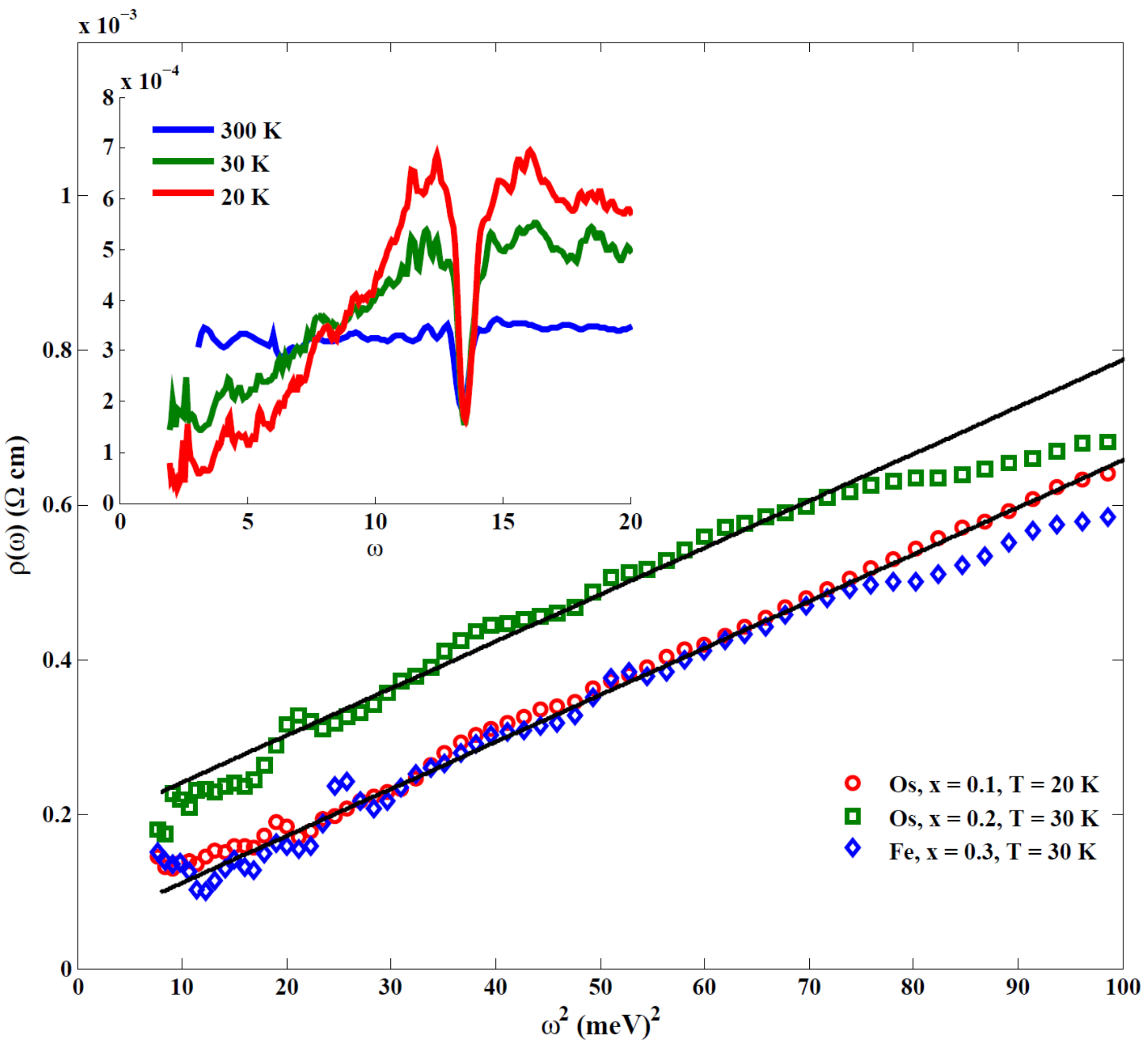}%
\caption{\label{rho_ac} The optical resistivity is linear when plotted against the square of the frequency. The closer to the transition the temperature at which $\rho(\omega)$ is measured, the higher in frequency the linear fit is valid. In all samples, regardless of the temperature of the transition or whether it was to an AFM or HO phase, the scattering is quadratic frequency immediately above the transition. The same behavior is seen in the parent compound, shown as an inset, which has linear scattering at 30 K that becomes quadratic closer to $T_0$. As the transition temperature rises with substitution, the temperature at which Fermi liquid behavior appears rises as well. }
\end{figure}

Figure \ref{rho_ac} shows the AC resistivity as a function of $\omega^{2}$ for the substituted compounds at temperatues just above the phase transition to the ordered state. It is linear in the low frequency regime  $\omega \lesssim$ 8.5 meV, indicating quadratic dependence of the scattering rate on frequency. This is true at temperatures well above the range where quadratic scattering rates are observed in the parent compound, indicating that \ursd\ is a coherent (though anomalous \cite{nagel12, movassagh14}, see below) Fermi liquid in the normal state in a narrow temperature range above the transition regardless of whether the transition is to the HO or the AFM phase. 

The unsubstituted compound is shown as an inset to Figure \ref{rho_ac}. At 30 K the scattering rate varies linearly with frequency, \textit{i.e.} it is non Fermi liquid like.  At 20 K coherence has developed with $1/\tau < \omega$ and the scattering rate varies as $\omega^2$. We show that with substitution the same behavior obtains: the ordered state, regardless of the order parameter, always emerges from a Fermi liquid precursor. This fact is not immediately apparent from transport measurements alone as the Fermi liquid temperature range is too narrow to establish a conventional $T^2$ dependence; it is only by looking at the optical scattering rate that it becomes clear that this must be the case.

For a Fermi liquid, in addition to the $\omega^2$ frequency dependence, one also expects a $T^2$ temperature dependence of the resistivity.  It was  shown that  the full resistivity is given by \cite{gurzhi59,maslov12}:
\begin{eqnarray}
\rho(\omega,T) = C(\omega^2 + b\pi^2 T^2)
\end{eqnarray}
where the value of the coefficient $C$ depends on the band structure but $b=4$ for umklapp scattering \cite{maslov12} independent of the details of a particular material. It was pointed out recently that for many strongly correlated systems the $b$ coefficient varied from 1.0 up to 2.5 and in particular for \urs\  it had a value of $b=1.0$ in a narrow range of temperatures above the hidden order transition \cite{nagel12}.  Maslov and Chubukov showed that this anomalous behavior can be the result of resonant elastic scattering \cite{maslov12}. In the case at hand the scattering centers would be the unhybridized $f$ electrons.

\begin{figure}
\hspace*{-0.6cm}\includegraphics[width=3.5 in]{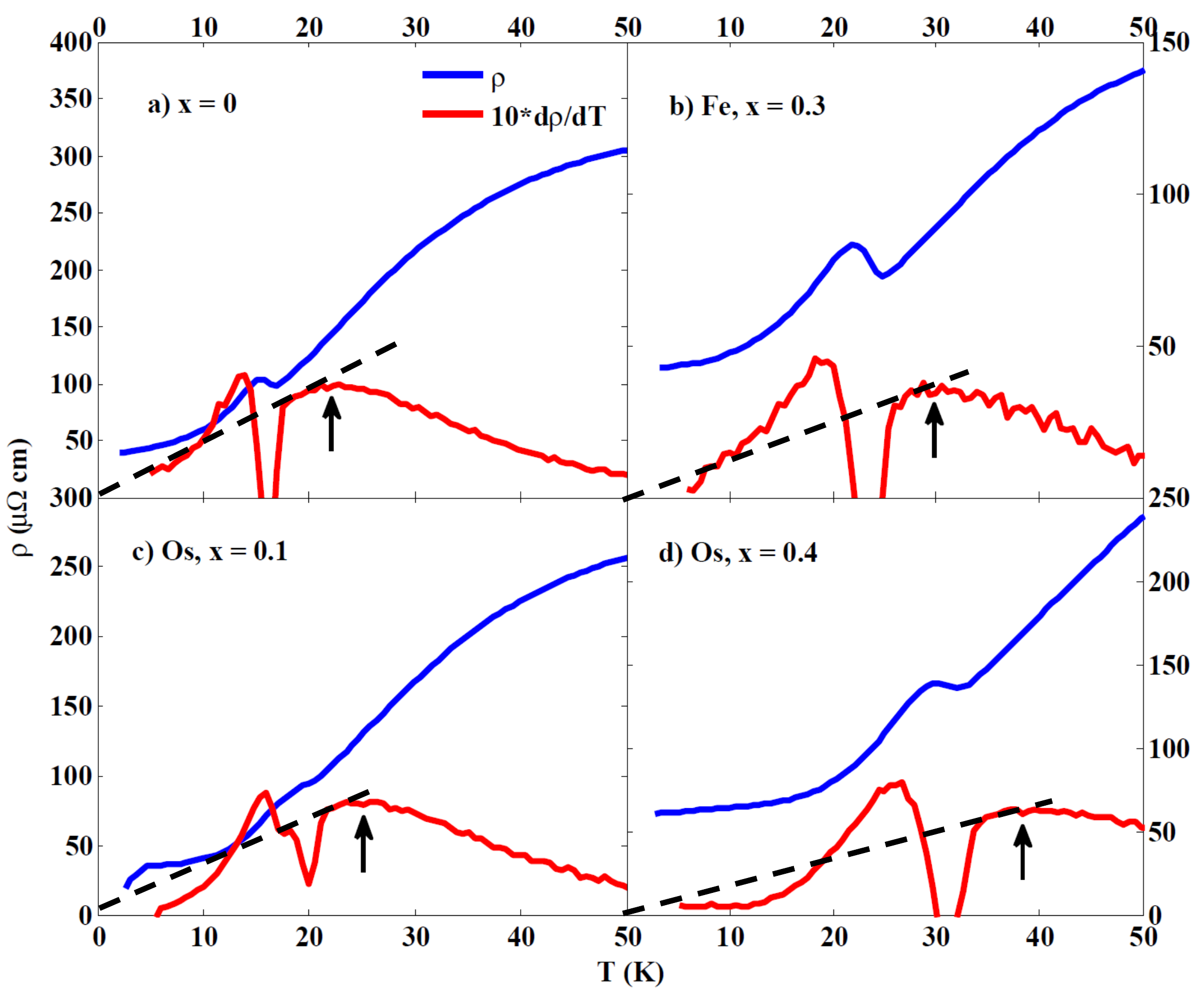}%
\caption{\label{rho_dc} The DC resistivity and its first derivative for \ursd.  The transition at $T_0$ to the ordered state is marked by a sudden sharp minimum in the resistivity. The resistivity of the parent compound has the same hallmarks as that of the substituted material. As the transition is approached from above the derivative reaches a maximum indicated by arrows in the figure, that always precedes the transition. The effect of substitution is to move the whole structure to higher temperatues preserving its overall features. The dashed line denotes Fermi liquid behavior where $d\rho(T)/dT=2AT$.}
\end{figure}

Figure \ref{rho_dc} shows the DC resistivity and its temperature derivative for \ursd. The resistivity bears many of the hallmarks of the resistivity of the parent compound \cite{nagel12}, with the transition marked by the same peak like feature that shifts up in temperature with $x$. The first derivative of the resistivity is characterized by a broad asymmetric peak above the transition. The transition itself is signaled as a sudden sharp drop in the derivative to negative values. With increasing $x$ this pattern shifts upwards in temperature. It is noteworthy that the turnover in the derivative has the same character in samples with a HO transition and those with an AFM transition. This is analogous to the behavior of the resistivity under pressure \cite{jeffries08}. 

Because the temperature range of Fermi liquid behavior is the narrow region between onset of coherence and the transition to the ordered state, we cannot use the usual method of plotting $\rho(T) = AT^2$ to determine the coefficient $A$ and then from it $b$. Instead we adopt the following procedure. Assuming that Eq. 1 holds we can determine $C$ from the slope of the frequency dependence as shown in Fig. \ref{rho_ac}. To find A we draw a straight line from the experimental resistivity derivative line to the origin as shown in Fig. \ref{rho_dc}b (dashed line), in effect assuming that $d\rho/dT=2AT$ where, in our notation, $A=Cb\pi^2$. Using this procedure we find that in the normal state at 19 K $b=1.1$ in the parent compound, while for the Fe $x=0.3$ material, just above the AFM transition at 30 K, $b=1.35$. So, in both materials, above the transition, there is a coherent Fermi liquid with anomalous $b$.   At the same temperature, in the parent material, the frequency dependence is not Fermi liquid like and the transport is incoherent, \textit{i.e.} $1/\tau > \omega$.  Thus we conclude that substitution with Fe and Os moves both the second order transition temperature and the region where we observe coherent $\omega^2$ Fermi liquid behavior in concert to higher temperatures. 

This observation indicates that regardless of the nature of the transition (AFM or HO) or the temperature at which it occurs, the dominant scattering mechanism for the charge carriers is due to scattering of coherent quasiparticles from resonant impurities. This was previously shown \cite{nagel12} to be the case for the parent material. We conclude that the emergence of this anomalous Fermi liquid scattering is a precondition for the occurrence of the ordered state.  Coherence and a well developed Femi surface are necessary conditions for a nesting induced density wave. What we have 
{\it not} observed in the normal state in \urs \ is scattering by discrete bosonic excitations; such excitations would be characterized by a distinct onset of scattering rather than the smooth $\omega^2$ dependence that we observe. (Such as for example the 41 meV magnetic resonance in the cuprates \cite{hwang07}).

In summary, we have studied two ordered phases of \urs\ spectroscopically: the hidden order phase and the antiferromagnetic phase.  The two phases show few differences other than an overall smooth increase in the gap and the transition temperature with substitution of Fe and Os.  In the ordered states the gap and the transfer of spectral weight are characteristic of density waves and consistent with a partial gapping scenario \cite{maple86}. The normal states are also very similar: they are Fermi liquid like with a scaling factor $b\approx1.0$ characteristic of a Fermi liquid dominated by resonant impurity scattering.  Models that include nesting-induced density waves are consistent with our observations: efficient nesting is promoted by coherent well-defined Fermi surfaces.

\begin{acknowledgments}
We thank P. Coleman, G. Kotliar, D.L. Maslov, S.S. Lee, B.D. Gaulin, and A.M. Tremblay for helpful discussions. 
This work was supported by the Natural Science and Engineering Research Council of Canada and by the Canadian Institute for Advanced Research.  The research at UCSD was supported by the U.S. Department of Energy, Office of Basic Energy Sciences, Division of Materials Sciences and Engineering under Grant No. DE-FG02-04ER46105 (sample synthesis) and the National Science Foundation under Grant No. DMR-1206553 (sample characterization). MJ acknowledges financial support by the Alexander von Humboldt foundation. 
\end{acknowledgments}

\bibliography{AFM_URu2Si2c}

\end{document}